
%
\input phyzzx
\catcode`@=11
%
%
\newtoks\UT
\newtoks\monthyear
\Pubnum={UT-\the\UT}
\UT={691}
\monthyear={November, 1994}
\def\p@bblock{\begingroup \tabskip=\hsize minus \hsize
    \baselineskip=1.5\ht\strutbox \topspace-2\baselineskip
    \halign to\hsize{\strut ##\hfil\tabskip=0pt\crcr
    \the\Pubnum\cr\the\monthyear\cr
    }\endgroup}
\def\bftitlestyle#1{\par\begingroup \titleparagraphs
    \iftwelv@\fourteenpoint\else\twelvepoint\fi
    \noindent{\bf #1}\par\endgroup}
\def\title#1{\vskip\frontpageskip \bftitlestyle{#1} \vskip\headskip}
%
%

%
%

%
\def\journal#1&#2(#3){\begingroup \let\journal=\dummyj@urnal
    \unskip, \sl #1\unskip~\bf\ignorespaces #2\rm
    (\afterassignment\j@ur \count255=#3) \endgroup\ignorespaces}
\def\andjournal#1&#2(#3){\begingroup \let\journal=\dummyj@urnal
    \sl #1\unskip~\bf\ignorespaces #2\rm
    (\afterassignment\j@ur \count255=#3) \endgroup\ignorespaces}
\def\andvol&#1(#2){\begingroup \let\journal=\dummyj@urnal
    \bf\ignorespaces #1\rm
    (\afterassignment\j@ur \count255=#2) \endgroup\ignorespaces}

\def\NP{Nucl.~Phys.}
\def\PL{Phys.~Lett.}
\def\PR{Phys.~Rev.}

\def\PTP{Prog.~Theor.~Phys.}

\catcode`@=12
%

\titlepage

\title{A Note on Color Confinement}

\author{Izawa {\twelverm Ken-Iti}
\foot{\rm JSPS Research Fellow.}}
\address{Department of Physics, University of Tokyo \break
                    Tokyo 113, Japan}

\abstract{
An argument is given which exhibits color confinement
in nonabelian gauge theory.
}

\endpage

\doublespace


\def\m{\mu}
\def\n{\nu}

\def\q{\partial}

\def\y{\eta}

\def\i{\int \!\!}
\def\o{\over}


\REF\Man{J.E.~Mandula and M.~Ogilvie \journal \PL &B185 (87) 127; \nextline
         C.~Bernard, C.~Parrinello, and A.~Soni \journal \PR &D49 (94) 1585.}

\REF\Kug{T.~Kugo and I.~Ojima \journal Suppl.~\PTP &66 (79) 1; \nextline
         N.~Nakanishi and I.~Ojima, {\sl Covariant Operator Formalism
         of Gauge Theories and Quantum Gravity} (World Scientific, 1990);
         \nextline
         H.~Hata and I.~Niigata \journal \NP &B389 (93) 133,
         and references therein.}

\sequentialequations

%
%
%

In this note, we point out that recent lattice study
\refmark{\Man}
on gluon propagator
supports Kugo-Ojima mechanism
\refmark{\Kug}
for color confinement
in nonabelian gauge theory.
It is assumed that color symmetry is not broken spontaneously.

The gluon propagator in the Landau gauge can be written
in momentum space as
$$
  \VEV{A_\m A_\n}
  = {1 \o i} (\y_{\m \n} - {p_\m p_\n \o p^2}) {1 \o p^2} R(p^2).
 \eqn\PROP
$$
Lattice calculations
\refmark{\Man}
suggest its nonsingular infrared behavior
$$
  \lim_{p^2 \rightarrow 0} R(p^2) = 0,
 \eqn\CO
$$
which implies that the field $A_\m$ in \PROP\
contains no massless one-particle
component of a genuine vector.

Then the term $\q^\m F_{\m \n}$ in the quantum Maxwell equation
$$
  \q^\m F_{\m \n} + g J_\n = \{Q_{_B}, D_\n {\bar c}\}
 \eqn\QME
$$
contains no massless one-particle component at all,
where $J_\n$ is the color current and $Q_{_B}$ denotes the BRS charge.
This is because a scalar one-particle component does not appear
in $\q^\m F_{\m \n}$ due to the antisymmetry of $F_{\m \n}$
in its Lorentz indices $\m$ and $\n$.
Under the assumption of unbroken color symmetry,
this indicates that the color charge is given by
$$
  Q = \{Q_{_B}, \i d{\bf x} \  g^{-1} D_0 {\bar c}\},
 \eqn\CC
$$
which is a manifestation of Kugo-Ojima mechanism
\refmark{\Kug}
for color confinement.

Thus we conclude that color confinement is realized in
nonabelian gauge theory
provided color symmetry is unbroken.


\endpage

\refout

\bye